\documentclass[12pt]{iopart}

\usepackage{cite}

\usepackage{graphicx}

\usepackage{dcolumn}

\begin{document}

\title[Exceptional points]{Exceptional points of the eigenvalues of parameter-dependent
Hamiltonian operators}

\author{Paolo Amore\dag \ and Francisco M Fern\'andez
\footnote[2]{Corresponding author}}

\address{\dag\ Facultad de Ciencias, Universidad de Colima, Bernal D\'iaz del
Castillo 340, Colima, Colima, Mexico}\ead{paolo@ucol.mx}

\address{\ddag\ INIFTA, Divisi\'on Qu\'imica Te\'orica,
Blvd. 113 S/N,  Sucursal 4, Casilla de Correo 16, 1900 La Plata,
Argentina}\ead{fernande@quimica.unlp.edu.ar}

\maketitle

\begin{abstract}
We calculate the exceptional points of the eigenvalues of several
parameter-dependent Hamiltonian operators of mathematical and physical
interest. We show that the calculation is greatly facilitated by the
application of the discriminant to the secular determinant. In this way the
problem reduces to finding the roots of a polynomial function of just one
variable, the parameter in the Hamiltonian operator. As illustrative
examples we consider a particle in a one-dimensional box with a polynomial
potential, the periodic Mathieu equation, the Stark effect in a polar rigid
rotor and in a polar symmetric top.
\end{abstract}

\section{Introduction}

In many applications of quantum mechanics to physical problems the
Hamiltonian operator $H(\lambda )$ depends on a parameter $\lambda $. For
example, in the case of atoms and molecules in electric or magnetic fields
the parameter $\lambda $ is related to the intensity of the external field
so that $H(0)$ is the Hamiltonian operator for the isolated atom or
molecule. One of the approximate methods for calculating the energies $%
E(\lambda )$ of such problems is perturbation theory that is based on the
Taylor expansion of the energies $E(\lambda )$ and eigenfunctions $\psi
(\lambda )$ about $\lambda =0$. The resulting series may be divergent or
they may have finite convergent radii\cite{RS78,S82,F01} (and references
therein).

When $H(\lambda )$ is an analytic function of $\lambda $ (the typical case
being $H(\lambda )=H_{0}+\lambda H^{\prime }$) the convergence of the
perturbation series radii are determined by exceptional points (EPS) in the
complex $\lambda $ plane where two or more eigenvalues coalesce. This
coalescence is different from degeneracy in that the corresponding
eigenvectors become linear dependent at an EP. For this reason there has
been great interest in the accurate calculation of EPS. Among the models
analyzed are the Mathieu equation\cite{BC69,H81,
HG81,FAC87,V98,SX99,ZRKBS12,FG14}, a polar rigid rotor in an electric field%
\cite{FC85,FG14}, a polar symmetric top in an electric field\cite%
{FAC87,MMFC87} and a particle in a box with a linear potential\cite{FG14}.
In all those cases only a pair of eigenvalues coalesce at each EP that is
also called a branch point, double point or critical point. The EPS on the
imaginary axis proved to be relevant to the study of $\mathcal{PT}$%
-symmetric non-Hermitian Hamiltonians\cite{FG14} (and references therein).

The EPS can be estimated by a suitable analysis of the perturbation series%
\cite{FAC87,F01} but there are more accurate techniques\cite{F01,BC69,H81,
HG81,V98,SX99,ZRKBS12,FG14}, most of which are based on the secular equation
for a truncated matrix representation of the Hamiltonian operator in a
suitable basis set of eigenvectors. In most of these cases, one obtains the
eigenvalues $E(\lambda )$ from the roots of a nonlinear equation $%
Q(E,\lambda )=0$ and it is well known that the branch points are
simultaneous solution of this equation and $\partial $ $Q(E,\lambda
)/\partial E=0$. This approach and its variants\cite{F01,BC69,H81,
HG81,V98,SX99,ZRKBS12,FG14} have proved suitable for the calculation of
reasonably accurate EPS. However, finding the roots of two nonlinear
equations requires a judicious application of efficient algorithms (see, for
example, the remarkable calculation of double points carried out many years
ago for the characteristic values of the Mathieu equation\cite{BC69}).

The purpose of this paper is to point out that the calculation of the EPS is
considerably facilitated by the application of the discriminant\cite%
{S1851,G81,KB16,BPR03} to the polynomial function that determines the
approximate energies of the quantum-mechanical problem. The advantage is
that the two nonlinear equations in two variables reduce to one nonlinear
equation in one variable. As a result, the estimation of the EPS reduces to
the calculation of the roots of one nonlinear equation in the parameter $%
\lambda $. Another advantage of this approach is that most computer-algebra
software have built-in algorithms for the calculation of the discriminant of
a polynomial. The resultant of two polynomials and the discriminant of a
polynomial are known since long ago in the mathematical literature\cite%
{S1851,G81,KB16,BPR03} and have already been applied to the analysis of
physical problems. Some of the examples are the determination of
singularities in the eigenvalues of parameter-dependent matrix eigenvalue
problems\cite{HS91}, the analysis of the properties of two-dimensional
magnetic traps for laser-cooled atoms\cite{D02}, the description of optical
polarization singularities\cite{F04}, the EPS for the eigenvalues of a
modified Lipkin model\cite{HSG05}, the location of level crossings between
eigenvalues of parameter-dependent symmetric matrices\cite%
{BR06,B07,BR07a,BR07b}, the solution of two equations with two unknowns that
appear in the study of gravitational lenses\cite{KB16}. In all these
applications the resultant and discriminant have been applied to polynomials
of finite degree coming from matrices of finite dimension where the
technique yields exact results. In this paper, on the other hand, we focus
on quantum-mechanical problems defined on infinite Hilbert spaces so that
the matrix representations of the Hamiltonian operators and their
characteristic polynomials are approximate due to necessary truncation.

In section~\ref{sec:Par-dep_Ham} we briefly discuss some properties of
parameter-dependent Hamiltonians, in section~\ref{sec:PB} we apply the
approach to a particle in a box with two different potentials, in section~%
\ref{sec:Mathieu} we consider the periodic solutions to the Mathieu
function, sections~\ref{sec:RR} and~\ref{sec:sym_top} are devoted to the
Stark effect in a polar rigid rotor and a symmetric top, respectively.
Finally, in section~\ref{sec:conclusions} we summarize the main results and
draw conclusions. In order to make this paper sufficiently self-contained we
add two appendices with a discussion of three-term recurrence relations for
the expansion coefficients of the wavefunctions and a slight introduction to
the resultant of two polynomials and the discriminant of a polynomial.

\section{Parameter-dependent Hamiltonians}

\label{sec:Par-dep_Ham}

The purpose of this paper is the analysis of Hamiltonian operators $%
H(\lambda )$ that depend on a parameter $\lambda $. Their eigenvalues $%
E_{n}(\lambda )$ and eigenvectors (or eigenfunctions) $\psi _{n}$ depend on
this parameter. This kind of problems may exhibit EPS $\lambda _{EP}$ in the
complex $\lambda $-plane at which two (or more) eigenvalues coalesce: $%
E_{m}\left( \lambda _{EP}\right) =E_{n}\left( \lambda _{EP}\right) $. This
phenomenon is different from ordinary degeneracy in that the corresponding
eigenvectors $\psi _{m}$ and $\psi _{n}$ become linearly dependent at an
exceptional point $\lambda _{EP}$. If we manage to obtain a power-series
expansion%
\begin{equation}
E_{n}(\lambda )=\sum_{j=0}E_{n,j}\lambda ^{j},
\label{eq:PTseriesE}
\end{equation}%
for example by means of perturbation theory, its radius of convergence will
be determined by the EP closest to the origin of the complex $\lambda $-plane%
\cite{RS78,S82,F01}.

One can obtain EPS from a nonlinear equation of the form $Q(E,\lambda )=0$,
from which one commonly obtains the eigenvalues $E_{n}(\lambda )$. It is
well known that the branch points of the eigenvalues as functions of the
complex parameter $\lambda $ are common roots of the nonlinear equations $%
\left\{ Q(E,\lambda )=0,\partial Q(E,\lambda )/\partial E=0\right\} $. In
this paper we consider a simple and straightforward way of obtaining the EPS
that enables one to reduce the two nonlinear equations of two variables to
just one nonlinear function of $\lambda $.

One of the simplest ways of obtaining the equation $Q(E,\lambda )=0$ is
based on the matrix representation $\mathbf{H}$ of the Hamiltonian operator $%
H$ in a given orthonormal basis set $\left\{ \left\vert i\right\rangle
,i=0,1,\ldots \right\} $. If the basis set is infinite we resort to an
approximate truncated matrix representation $\mathbf{H}_{N}$ with elements $%
H_{ij}=\left\langle i\right\vert H\left\vert j\right\rangle $, $%
i,j=0,1,\ldots ,N-1$. The approximate energies are roots of the
characteristic polynomial
\begin{equation}
p_{N}(E,\lambda )=\left\vert
\mathbf{H}_{N}-E\mathbf{I}_{N}\right\vert =0,
\label{eq:p_N(E,lambda)}
\end{equation}%
where $\mathbf{I}_{N}$ is the $N\times N$ identity matrix. The roots of $%
p_{N}(E,\lambda )=0$ give us approximate eigenvalues $E_{n}(\lambda )$ and,
consequently, we expect to obtain the EPS with increasing accuracy by
increasing $N$.

Since the EPS are complex values of $\lambda $ for which two
eigenvalues coalesce, and taking into account that
$p_{N}(E,\lambda )$ is a polynomial function of $E$, it is clear
from the discussion of the discriminant in~\ref{app:Discriminant}
that the EPS can be obtained from the roots $\lambda _{i}^{[N]}$
of the one-variable function%
\begin{equation}
F_{N}(\lambda )=Disc_{E}\left( p_{N}(E,\lambda )\right).
\label{eqF_N=Disc(p_N)}
\end{equation}%
Since the characteristic polynomial comes from a truncation of the matrix
representation of the Hamiltonian, one expects some of the roots of $%
F_{N}(\lambda )$ to be spurious. However, the sequences of roots
that converge when $N$ increases are expected to yield the actual
EPS. Another advantage of this approach is that most
computer-algebra software have built-in algorithms for the
calculation of the discriminant of a polynomial. It is worth
noticing that this approach applies even in the case that more
than two eigenvalues coalesce at an EP
(see~\ref{app:Discriminant}). The approach just outlined is
particularly simple and practical when $H=H_{0}+\lambda H^{\prime
}$ because in this case $F_{N}(\lambda )$ is a polynomial function
of $\lambda $.

Suppose that $E(0)$ is an isolated simple eigenvalue of $H_{0}$ and that
there are two real numbers $a$ and $b$ such that%
\begin{equation}
\left\Vert H^{\prime }\Phi \right\Vert \leq a\left\Vert H_{0}\Phi
\right\Vert +b\left\Vert \Phi \right\Vert,
\label{eq:Kato_inequality}
\end{equation}%
where $\left\Vert f\right\Vert =\sqrt{\left\langle f\right\vert \left.
f\right\rangle }$, for all $\Phi $ in the state space. Under such conditions
there is a unique eigenvalue $E(\lambda )$ of $H$ near $E(0)$ and $E(\lambda
)$ is analytic in a neighbourhood of $\lambda =0$ in the complex $\lambda $
plane\cite{RS78,S82,F01} (and references therein). All the examples
discussed in this paper satisfy this condition because $\left\Vert H^{\prime
}\Phi \right\Vert \leq b\left\Vert \Phi \right\Vert $ as we will see below.

If $H(\lambda )$ is Hermitian for $\lambda $ real, then the EPS cannot be
real. In all the examples discussed here $H(\lambda )^{\ast }=H(\lambda
^{\ast })$ for $\lambda $ complex so that $H(\lambda ^{\ast })\psi
_{n}^{\ast }=E_{n}(\lambda )^{\ast }\psi _{n}^{\ast }$ and, consequently, $%
E_{n}(\lambda )^{\ast }=E_{m}(\lambda ^{\ast })$. Suppose that $%
E_{n}(\lambda )$ and $E_{k}(\lambda )$ coalesce at $\lambda _{EP}$,
therefore $E_{n}\left( \lambda _{EP}\right) ^{\ast }=E_{k}\left( \lambda
_{EP}\right) ^{\ast }\Rightarrow E_{m}\left( \lambda _{EP}^{\ast }\right)
=E_{j}\left( \lambda _{EP}^{\ast }\right) $ and, by virtue of the preceding
result, $\lambda _{EP}^{\ast }$ is also an EP. In other words, the EPS
appear in pairs of complex-conjugate numbers.

Suppose that there exists a unitary operator $U$ such that $U^{\dagger
}H(\lambda )U=H(-\lambda )$; therefore $U^{\dagger }H(\lambda )\psi
_{n}=H(-\lambda )U^{\dagger }\psi _{n}=E_{n}(\lambda )U^{\dagger }\psi _{n}$%
, from which it follows that $E_{n}(\lambda )=E_{m}(-\lambda )$. We conclude
that if $\lambda _{EP}$ is an EP, then $-\lambda _{EP}$ is also an EP. It
follows from the two preceding results that there may be quadruplets of EPS:
$\lambda _{EP}$, $\lambda _{EP}^{\ast }$, $-\lambda _{EP}$ and $-\lambda
_{EP}^{\ast }$ when $\Re \lambda _{EP}\neq 0$.

\section{Particle in a one-dimensional box with a perturbation}

\label{sec:PB}

Our first example is given by a particle in a one-dimensional box with an
interaction potential. The Schr\"{o}dinger equation for such a problem can
be written in dimensionless form as%
\begin{eqnarray}
&&-\psi ^{\prime \prime }(x)+\lambda V(x)\psi (x)=E\psi (x),  \nonumber \\
&&\psi (-1) =\psi(1)=0.  \label{eq:PB_V}
\end{eqnarray}%
If $V(x)=H^{\prime }$ is a continuous function of $x$ then $\left\vert
V(x)\right\vert \leq C$ when $x\in (-1,1)$ and $\left\Vert H^{\prime }\Phi
\right\Vert \leq C^{2}\left\Vert \Phi \right\Vert $. Consequently, every
eigenvalue $E_{n}(\lambda )$ is an analytical function of $\lambda $ for all
$\lambda <\left\vert \lambda _{EP}\right\vert $, where $\lambda _{EP}$ is
the EP closest to the origin of the complex $\lambda $ plane. In order to
obtain a suitable matrix representation for this Hamiltonian we resort to
the basis set of eigenfunctions of $H_{0}=-d^{2}/dx^{2}$:%
\begin{equation}
\left\langle x\right. \left\vert n\right\rangle =\sin \left(
\frac{(n+1)\pi (x+1)}{2}\right) ,\;n=0,1,\ldots .
\label{eq:PB_basis}
\end{equation}

If $V(x)$ is a polynomial function of $x$ the analytical calculation of its
matrix elements $\left\langle m\right\vert V\left\vert n\right\rangle $ is
straightforward. The simplest example is $V(x)=x$. The canonical
transformation $U^{\dagger }xU=-x$, $U^{\dagger }d/dxU=-d/dx$ yields $%
U^{\dagger }H(\lambda )U=H(-\lambda )$ and, therefore, we expect the EPS to
appear as quadruplets ($\Re \lambda _{EP}\neq 0$) and doublets ($\Re \lambda
_{EP}=0$).

\begin{figure}[tbp]
\begin{center}
\includegraphics[width=9cm]{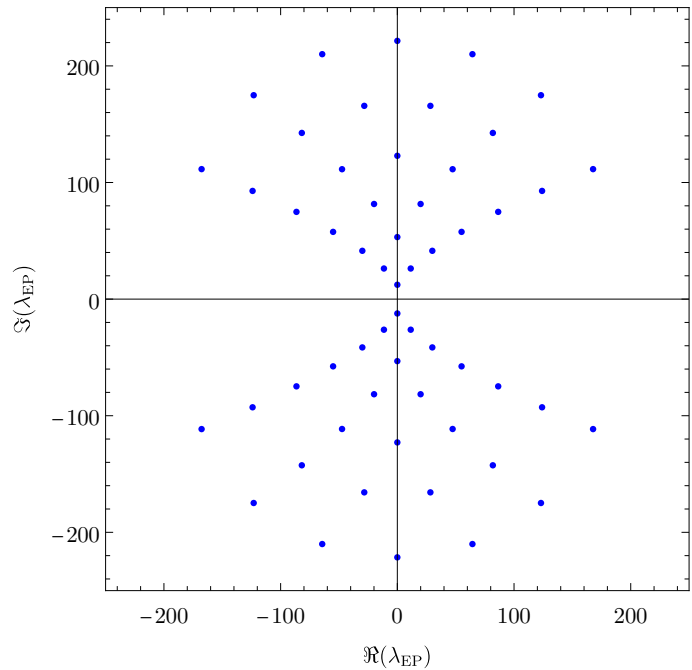}
\end{center}
\caption{EPS for the particle in a box with potential $\protect\lambda x$}
\label{fig:PBX1EPS}
\end{figure}

In order to discard any spurious root $\lambda _{i}^{[N]}$ of the polynomial
$F_{N}(\lambda )$ of degree $N(N-1)$ we keep only those that satisfy $%
\left\vert \lambda _{EP}^{[N]}-\lambda _{EP}^{[N-1]}\right\vert <10^{-3}$.
Figure~\ref{fig:PBX1EPS} shows some of the EPS for this problem; they
clearly exhibit the symmetry just mentioned with respect to the real ($%
\lambda _{EP}$,$\lambda _{EP}^{\ast }$) and imaginary ($\lambda _{EP}$,$%
-\lambda _{EP}$) axes of the complex $\lambda $ plane. A larger number of
EPS on the imaginary axis was obtained recently in a study of $\mathcal{PT}$%
-symmetric non-Hermitian Hamiltonians\cite{FG14}.

\begin{figure}[tbp]
\begin{center}
\includegraphics[width=9cm]{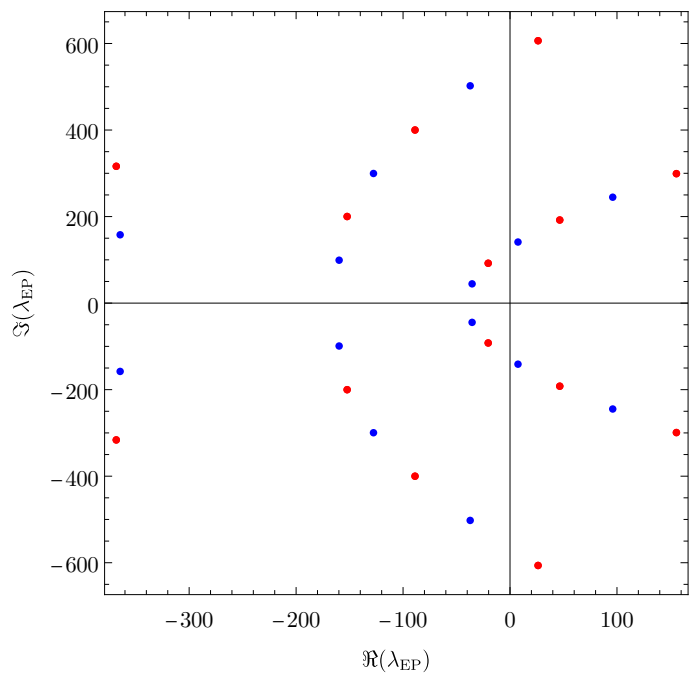}
\end{center}
\caption{EPS for the particle in a box with potential $\protect\lambda x^{2}$
for even (blue circles) and odd (red circles) states}
\label{fig:PBX2EPS}
\end{figure}

If $V(x)=x^{2}$ the canonical transformation discussed above leaves $%
H(\lambda )$ invariant and we only expect doublets (that is to say: symmetry
with respect to the real axis). Figure~\ref{fig:PBX2EPS} confirms this
conclusion. In this case it is convenient to treat the even ($n=0,2,\ldots $%
) and odd states ($n=1,3,\ldots $) separately.

\section{Mathieu equation}

\label{sec:Mathieu}

One of the most widely studied periodic problems is the Mathieu equation
that we write here in the following form%
\begin{equation}
\psi ^{\prime \prime }(x)+\left[ E-2\lambda \cos (2x)\right] \psi
(x)=0, \label{eq:Mathieu}
\end{equation}%
so that we can relate it to the linear operator
$H=-d^{2}/dx^{2}+2\lambda \cos (2x)$. We consider the two cases of
periodic solutions, those of period $\pi $ ($\psi (x+\pi )=\psi
(x)$) and those of period $2\pi $ ($\psi (x+2\pi )=\psi (x)$) and
each class can be separated into even and odd. The four cases can
be reduced to tridiagonal matrix representations or three-term
recurrence relations; in what follows we show the main parameters
(see~\ref{app:Rec-Rel}) for each of them.

Period $\pi $ even:
\begin{eqnarray}
\left\vert j\right\rangle &=&\frac{\sqrt{2}+\left( 1-\sqrt{2}\right) \delta
_{j0}}{\sqrt{\pi }}\cos (2jx),\;j=0,1,\ldots,  \nonumber \\
A_{j+1} &=&\left[ 1+\left( \sqrt{2}-1\right) \delta _{j0}\right]
\lambda ,\;B_{j}=4j^{2}-E.  \label{eq:Mathieu_pi_even}
\end{eqnarray}

Period $\pi $ odd%
\begin{eqnarray}
\left\vert j\right\rangle &=&\sqrt{\frac{2}{\pi }}\sin
[(2j+2)x],\;j=0,1,\ldots,  \nonumber \\
A_{j+1} &=&\lambda ,\;B_{j}=4(j+1)^{2}-E.
\label{eq:Mathieu_pi_odd}
\end{eqnarray}

Period $2\pi $ even%
\begin{eqnarray}
\left\vert j\right\rangle &=&\sqrt{\frac{2}{\pi }}\cos \left[ (2j+1)x\right]
,\;j=0,1,\ldots,  \nonumber \\
A_{j+1} &=&\lambda ,\;B_{j}=(2j+1)^{2}+\lambda \delta _{j0}-E.
\label{eq:Mathieu_2pi_even}
\end{eqnarray}

Period $2\pi $ odd%
\begin{eqnarray}
\left\vert j\right\rangle &=&\sqrt{\frac{2}{\pi }}\sin \left[ (2j+1)x\right]
,\;j=0,1,\ldots,  \nonumber \\
A_{j+1} &=&\lambda ,\;B_{j}=(2j+1)^{2}-\lambda \delta _{j0}-E.
\label{eq:Mathieu_2pi_odd}
\end{eqnarray}

In all these cases we find that $\left\Vert H^{\prime }\Phi \right\Vert \leq
\left\Vert H^{\prime }\Phi \right\Vert $ because $\left\vert H^{\prime
}\right\vert =\left\vert \cos (2x)\right\vert \leq 1$. Besides, the
canonical transformation $U^{\dagger }xU=x+\pi /2$, $U^{\dagger }d/dxU=d/dx$
leads to $U^{\dagger }H(\lambda )U=H(-\lambda )$ and again we expect the
distribution of EPS to exhibit symmetry with respect to both axes.

\begin{figure}[tbp]
\begin{center}
\includegraphics[width=9cm]{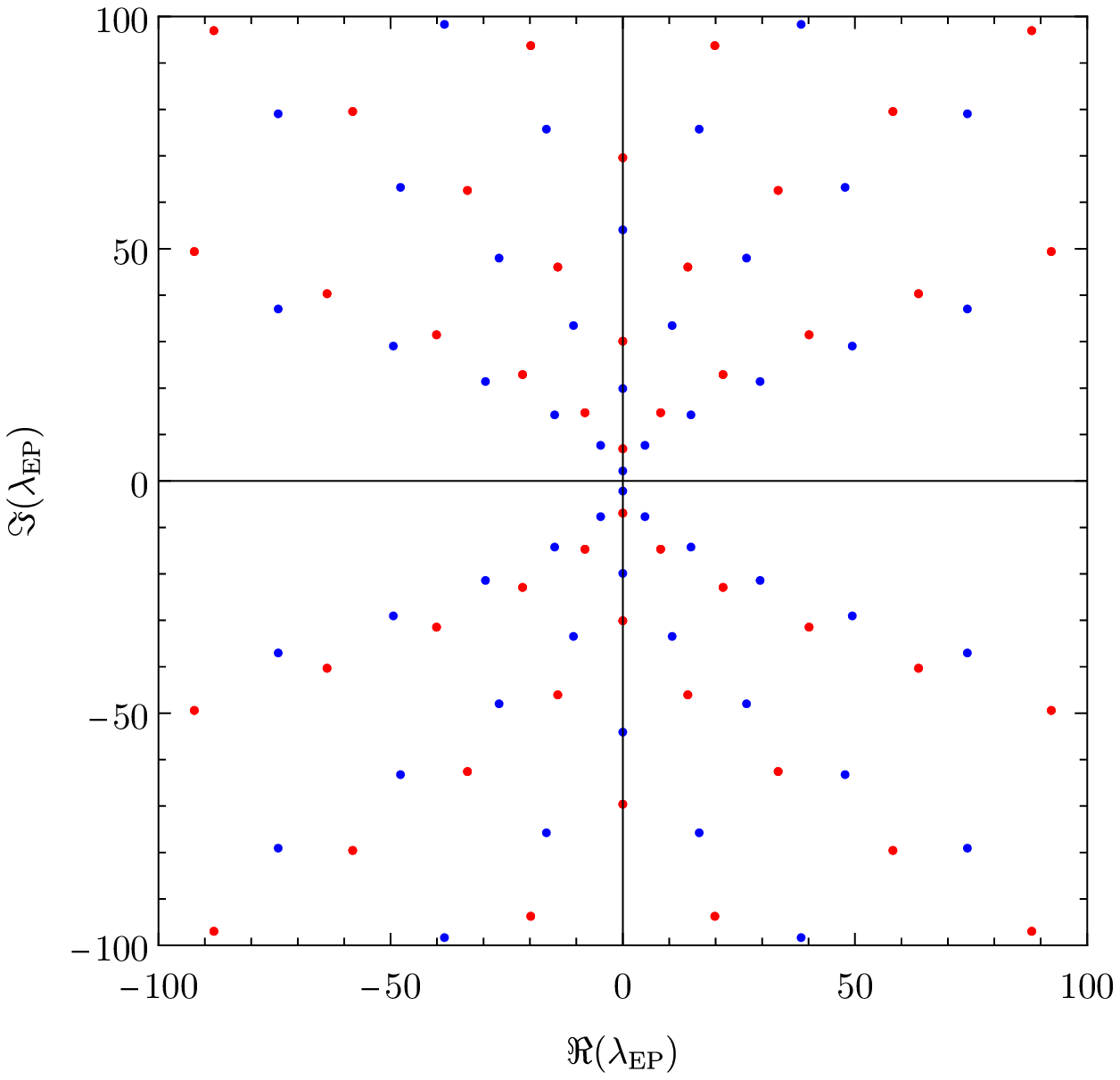}
\end{center}
\caption{EPS for the Mathieu equation with period $\protect\pi$ for even
(blue circles) and odd (red circles) functions}
\label{fig:Mathieupi}
\end{figure}

\begin{figure}[tbp]
\begin{center}
\includegraphics[width=9cm]{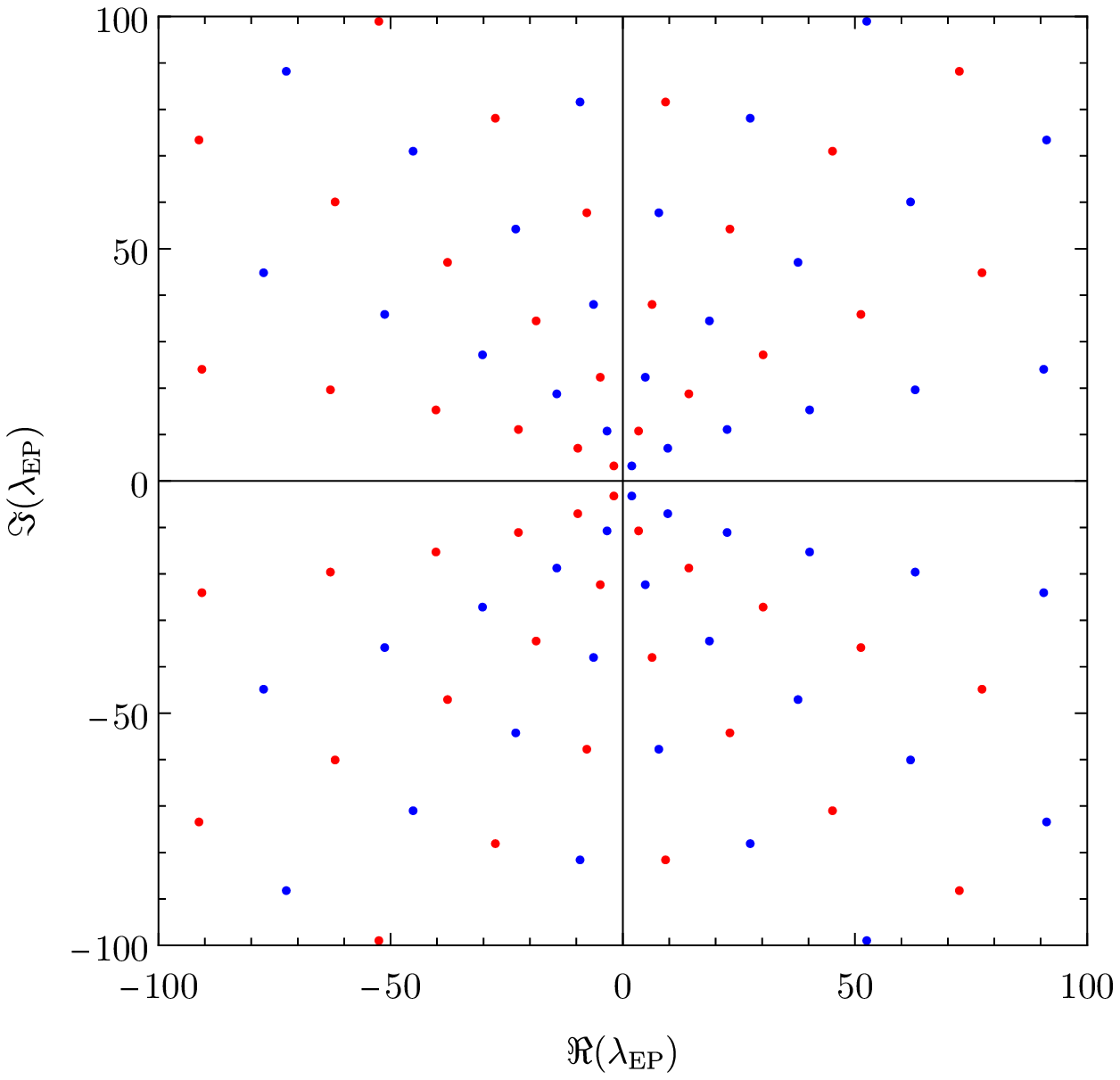}
\end{center}
\caption{EPS for the Mathieu equation with period $2\protect\pi$ for even
(blue circles) and odd (red circles) functions}
\label{fig:Mathieu2pi}
\end{figure}

Since in the four cases we have tridiagonal matrices, we resort to
the recurrence relation for the determinants $D_{N}$ discussed
in~\ref{app:Rec-Rel}. Thus, the problem reduces to obtaining the
roots of a polynomial $F_N(\lambda )=Disc_{E}(D_{N}(E,\lambda ))$
of degree $N(N+1)$. Present results are shown in figures
\ref{fig:Mathieupi} and \ref{fig:Mathieu2pi}. In the case of
period $\pi $ the distribution of EPS for each parity symmetry
(even and odd) exhibit the characteristic symmetry with respect to
both axes. However,
in the case of those of period $2\pi $ the even functions exhibit EPS $%
\lambda _{EP}$ and $\lambda _{EP}^{\ast }$ while the odd functions exhibit
the remaining ones $-\lambda _{EP}$ and $-\lambda _{EP}^{\ast }$. The reason
is that for period $\pi $ $D_{N}^{e,o}(E,-\lambda )=D_{N}^{e,o}(E,\lambda )$
while for period $2\pi $ $D_{N}^{e}(E,-\lambda )=D_{N}^{o}(E,\lambda )$,
where the superscripts $e$ and $o$ stand for even and odd parity,
respectively.

Present results agree with those of Blanch and Clemm\cite{BC69} in the whole
$\lambda $ plane and the ones obtained by Fern\'{a}ndez and Garcia\cite{FG14}
on the imaginary axis. Both appear to be the most accurate EPS available in
the literature.

\section{Polar rigid rotor in a uniform electric field}

\label{sec:RR}

In this section we consider a rigid rotor with dipole moment $\mu $ in a
uniform electric field of intensity $F$. The kinetic energy of the
Hamiltonian operator is $\mathcal{L}^{2}/(2I)$, where, $\mathcal{L}^{2}$ is
the square of the angular momentum and $I$ is the moment of inertia. The
interaction with the field is $-\mu F\cos \theta $, where $\theta $ is the
angle between the dipole moment and the field direction. This model has
proved useful for the analysis of the rotational Stark effect in linear
polar molecules\cite{TS55}. The Schr\"{o}dinger equation can be written in
dimensionless form as $H(\lambda )\psi =\epsilon (\lambda )\psi $, where%
\begin{equation}
H(\lambda )=H_{0}+\lambda H^{\prime }=L^{2}-\lambda \cos \theta,
\label{eq:Rigid_Rotor}
\end{equation}%
$L^{2}=\mathcal{L}^{2}/\hbar ^{2}$, $\lambda =2I\mu F/\hbar ^{2}$ and $%
\epsilon =2IE/\hbar ^{2}$. As in the preceding example we have $\left\Vert
H^{\prime }\Phi \right\Vert \leq \left\Vert \Phi \right\Vert $ and because
of the transformation $U^{\dagger }\theta U=\theta +\pi \Rightarrow
U^{\dagger }H(\lambda )U=H(-\lambda )$ we expect a distribution of EPS that
is symmetric with respect to both axes in the complex $\lambda $ plane.

In order to apply present approach we resort to a matrix representation of
the Hamiltonian operator in the basis set of eigenfunctions of $L^{2}$ and $%
L_{z}$%
\begin{eqnarray}
L^{2}\left\vert l,m\right\rangle &=&l(l+1)\left\vert l,m\right\rangle
,\;l=0,1,\ldots ,  \nonumber \\
L_{z}\left\vert l,m\right\rangle &=&m\left\vert l,m\right\rangle
,\;m=0,\pm 1,\ldots ,\pm l.  \label{eq:Angular_momentum_basis}
\end{eqnarray}%
It is well known that the coefficients $c_{i}$ of the expansion%
\begin{equation}
\psi =\sum_{i=0}^{\infty }c_{i}\left\vert M+i,m\right\rangle
,\;M=|m|, \label{eq:psi_RR_expansion}
\end{equation}%
satisfy a three-term recurrence relation like those discussed in the~\ref{app:Rec-Rel} with\cite{TS55,F01}%
\begin{equation}
A_{i}=-\lambda \left[ \frac{i(i+2M)}{4(i+M)^{2}-1}\right] ^{1/2},%
\;B_{i}=(i+M)(i+M+1)-\epsilon.  \label{eq:A_i,B_i_RR}
\end{equation}

\begin{figure}[tbp]
\begin{center}
\includegraphics[width=9cm]{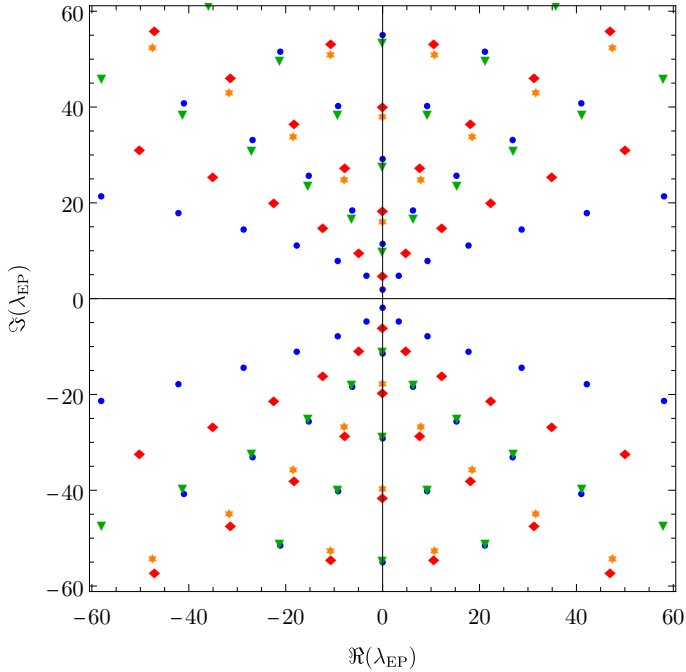}
\end{center}
\caption{EPS for the rigid rotor in an electric field for the states with $%
M=0$ (blue circles) $M=1$ (red diamonds) $M=2$ (green triangles) and $M=3$
(orange stars)}
\label{fig:RREPS}
\end{figure}

The calculation of the EPS $\lambda _{EP}$ from the determinants $%
D_{N}(\epsilon ,\lambda )=0$ (see~\ref{app:Rec-Rel}) is
straightforward and some results are shown in
figure~\ref{fig:RREPS} for $M=0,1,2,3$. It is worth noticing that
the EPS for $M=0$ are close to those for $M=2$ while the EPS for
$M=1$ are close to those for $M=3$. Notice that the distribution
of EPS exhibits the symmetry with respect to both axes mentioned
above.

\section{Polar symmetric top in a uniform electric field}

\label{sec:sym_top}

The rotational Stark effect in a symmetric-top molecule is commonly studied
by means of the model Hamiltonian\cite{TS55,S63,HO91}%
\begin{eqnarray}
H &=&-\frac{\hbar ^{2}}{2I_{B}}\left[ \frac{1}{\sin \theta }\frac{\partial }{%
\partial \theta }\sin \theta \frac{\partial }{\partial \theta }+\frac{1}{%
\sin ^{2}\theta }\frac{\partial ^{2}}{\partial \phi ^{2}}+\left( \frac{\cos
^{2}\theta }{\sin ^{2}\theta }+\frac{I_{B}}{I_{C}}\right) \frac{\partial ^{2}%
}{\partial \chi ^{2}}-\frac{2\cos \theta }{\sin ^{2}\theta }\frac{\partial
^{2}}{\partial \phi \partial \chi }\right]  \nonumber \\
&&-\mu F\cos \theta,  \label{eq:symmetric_top_1}
\end{eqnarray}%
where $I_{C}$ is the moment of inertia about the symmetry axis, $I_{B}$ is
the other moment of inertia, $\phi $, $\theta $ and $\chi $ are the Euler
angles, $\mu $ is the dipole moment of the molecule and $F$ is the intensity
of the uniform electric field. Clearly, $\theta $ is the angle between the
dipole moment and the field direction.

The Schr\"{o}dinger equation $H\psi =E\psi $ is separable if we write $\psi
(\theta ,\phi ,\chi )=\Theta (\theta )e^{iM\phi }e^{iK\chi }$, where $%
M,K=0,\pm 1,\pm 2,\ldots $ are the two rotational quantum numbers that
remain when the field is applied. The resulting eigenvalue equation in
dimensionless form is%
\begin{eqnarray}
&&\left( \frac{1}{\sin \theta }\frac{d}{d\theta }\sin \theta \frac{d}{%
d\theta }-\frac{M^{2}}{\sin ^{2}\theta }-\frac{\cos ^{2}\theta }{\sin
^{2}\theta }K^{2}+\frac{2\cos \theta }{\sin ^{2}\theta }KM-\lambda \cos
\theta +\epsilon \right) \Theta =0,  \nonumber \\
&&\epsilon =\frac{2I_{B}}{\hbar ^{2}}E-\frac{I_{B}}{I_{C}}K^{2},\;\lambda =%
\frac{2I_{B}}{\hbar ^{2}}\mu F.  \label{eq:symmetric_top_2}
\end{eqnarray}

If we write $H(\lambda )=H_{0}+\lambda H^{\prime }$ then we realize that $%
\left\Vert H^{\prime }\Phi \right\Vert \leq \left\Vert \Phi \right\Vert $ as
in the two preceding examples. Besides, the transformation used in the case
of the rigid rotor leads to $U^{\dagger }H(\lambda )U=H(-\lambda )$ and once
again we expect a distribution of EPS that is symmetric with respect to both
axes.

In this case we obtain a suitable matrix representation of the Hamiltonian
in the basis set of eigenvectors $\left\vert J,M,K\right\rangle $ of the
free symmetric top ($\lambda =0$). The coefficients $c_{i}$ of the expansion
\begin{equation}
\psi =\sum_{i=0}^{\infty }c_{i}\left\vert J_{0}+i,M,K\right\rangle
,\;J_{0}=\max (|M|,|K|),  \label{eq:psi_symmetric_top_expansion}
\end{equation}%
satisfy a three-term recurrence relation with\cite{TS55,S63,HO91}%
\begin{eqnarray}
B_{i} &=&\left( J_{0}+i\right) \left( J_{0}+i+1\right) -\lambda \frac{MK}{%
\left( J_{0}+i\right) \left( J_{0}+i+1\right) }-\epsilon,  \nonumber \\
A_{i} &=&-\lambda \frac{\sqrt{\left[ \left( J_{0}+i\right) ^{2}-K^{2}\right] %
\left[ \left( J_{0}+i\right) ^{2}-M^{2}\right] }}{\left(
J_{0}+i\right) \sqrt{4\left( J_{0}+i\right) ^{2}-1}}.
\label{eq:A_i,B_i_symmetric_top}
\end{eqnarray}%
Therefore, we can obtain the EPS from the secular determinants $%
D_{N}(\epsilon ,\lambda )=0$ for a sufficiently large dimension
$N$ (see~\ref{app:Rec-Rel}).

\begin{figure}[tbp]
\begin{center}
\includegraphics[width=9cm]{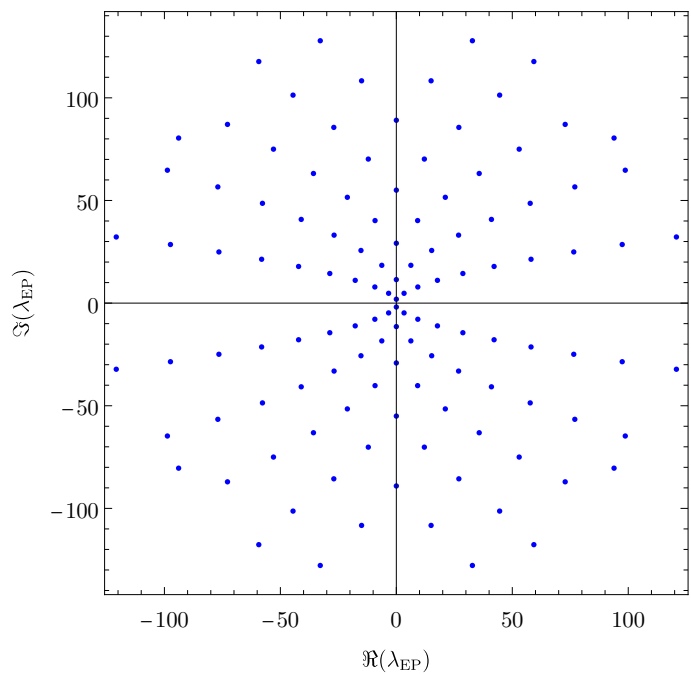}
\end{center}
\caption{EPS for the polar symmetric top in a uniform electric field for $%
M=0 $ and $K=0$}
\label{fig:SYMTOPM0K0}
\end{figure}

The distribution of the EPS can be predicted from the set of equalities%
\begin{eqnarray}
&&D_{N}(\epsilon ,M,K,\lambda )=D_{N}(\epsilon ,K,M,\lambda
)=D_{N}(\epsilon ,-M,-K,\lambda )=D_{N}(\epsilon ,-K,-M,\lambda )=
\nonumber
\\
&&D_{N}(\epsilon ,-M,K,-\lambda ) =D_{N}(\epsilon ,K,-M,-\lambda
)=D_{N}(\epsilon ,M,-K,-\lambda )=\nonumber \\  &&D_{N}(\epsilon
,-K,M,-\lambda ). \label{eq:D_N_symmetric_top_symmetry}
\end{eqnarray}%
Figure \ref{fig:SYMTOPM0K0} shows that the distribution of EPS for $M=K=0$
is symmetric with respect to both axes. On the other hand, Figure \ref%
{fig:SYMTOPM1K1M1KM1} shows that the distribution of EPS for either $MK=1$
or $MK=-1$ is symmetric with respect to the real axis but the union of both
sets is symmetric with respect to both axes.

Present results are considerably more accurate than those obtained earlier
for this model\cite{FAC87,MMFC87}.

\section{Conclusions}

\label{sec:conclusions}

In this paper we have shown that the discriminant is an extremely useful
tool for the location of EPS in the eigenvalues of parameter-dependent
Hamiltonian operators. In all the previous studies that we are aware of, the
approach was applied to operators on finite Hilbert spaces with matrix
representations of finite dimension that lead to characteristic polynomials
of finite degree\cite{SM98,HS91,D02,F04,HSG05,BR06,B07,BR07a,BR07b,KB16}.
Here, on the other hand, the Hilbert spaces have infinite dimension so that
the truncated $N$-dimensional matrix representations, as well as the
corresponding characteristic polynomials are approximate. However, the
location approach based on the discriminant applies successfully producing
sequences of roots that converge towards the actual EPS as $N$ increases.
Any spurious root is easily identified because it does not form part of a
convergent sequence.

All the examples chosen in the present study are of mathematical or physical
interest. Our results for the Mathieu equation agree with the most extended
and accurate ones available in the literature\cite{BC69} and those for the
Stark effects in the rigid rotor and symmetric top are either more extended
or more accurate than the ones published previously\cite%
{F01,FAC87,FG14,FC85,MMFC87}.

\ack{The research of P.A. was supported by Sistema nacional de
Investigadores (M\'exico). F.M.F. acknowledges a travel grant from
the University of La Plata.}

\appendix

\section{Resultant and discriminant}

\label{app:Discriminant}

In this section we summarize those properties of the discriminant of a
polynomial that are relevant for present paper. The resultant of two
polynomials%
\begin{eqnarray}
f(x) &=&\sum_{j=0}^{m}a_{m-j}x^{j},  \nonumber \\
g(x) &=&\sum_{j=0}^{n}b_{m-j}x^{j},  \label{eq:polynomials}
\end{eqnarray}%
is given by the determinant%
\begin{equation}
Res_{x}(f,g)=\left\vert
\begin{array}{cccccccc}
a_{0} & 0 & \cdots  & 0 & b_{0} & 0 & \cdots  & 0 \\
a_{1} & a_{0} & \cdots  & 0 & b_{1} & b_{0} & \cdots  & 0 \\
a_{2} & a_{1} & \ddots  & 0 & b_{2} & b_{1} & \ddots  & \vdots  \\
\vdots  & \vdots  & \ddots  & a_{0} & \vdots  & \vdots  & \ddots  & b_{0} \\
a_{m} & a_{m-1} & \cdots  & \vdots  & b_{n} & b_{n-1} & \cdots  & \vdots  \\
0 & a_{m} & \ddots  & \vdots  & 0 & b_{n} & \ddots  & \vdots  \\
\vdots  & \vdots  & \ddots  & a_{m-1} & \vdots  & \vdots  & \ddots  & b_{n-1}
\\
0 & 0 & \cdots  & a_{m} & 0 & 0 & \cdots  & b_{n}%
\end{array}%
\right\vert  . \label{eq:Res(f,g)_1}
\end{equation}%
It can be proved that%
\begin{equation}
Res_{x}(f,g)=a_{0}^{m}b_{0}^{n}\prod_{i=1}^{m}\prod_{j=1}^{n}\left(
\xi _{i}-\mu _{j}\right),   \label{eq:Res(f,g)_2}
\end{equation}%
where $\xi _{i}$ and $\mu _{j}$ are the roots of the polynomials $f$ and $g$%
, respectively. The discriminant of $f(x)$ is defined as%
\begin{equation}
Disc_{x}(f)=\frac{(-1)^{m(m-1)/2}}{a_{m}}Res_{x}(f,f^{\prime }),
\label{eq:Disc(f)_1}
\end{equation}%
and in this case we have%
\begin{equation}
Disc_{x}(f)=a^{2m-2}\prod_{i<j}\left( \xi _{i}-\xi _{j}\right)
^{2}. \label{eq:Disc(f)_2}
\end{equation}

Suppose that the nonlinear equation $Q(E,\lambda )=0$ gives us the
eigenvalues $E(\lambda )$ of a quantum-mechanical system. If this equation
is a polynomial function of $E$, then the roots of $F(\lambda
)=Disc_{E}(Q(E,\lambda ))$ are the exceptional points $\lambda _{EP}$ in the
complex $\lambda $ plane where at least two eigenvalues coalesce. We
appreciate that the advantage of resorting to the discriminant is that we
only have to search for the roots of a nonlinear function of just one
variable. In all the examples studied here the nonlinear equation $%
Q(E,\lambda )$ is a polynomial function of both $E$ and $\lambda $ so that $%
F(\lambda )$ is a polynomial function of $\lambda $
(see~\ref{app:Rec-Rel}, and the examples). Consequently, the
calculation is particularly simple because there are efficient
algorithms for finding the roots of polynomials. Besides, most
computer-algebra software enable one to obtain analytical
expressions for $F(\lambda )$ because the discriminant is given by
a determinant. Thus, the only numerical step of the calculation
reduces to finding the roots of the polynomial $F(\lambda )$.

\begin{figure}[tbp]
\begin{center}
\includegraphics[width=9cm]{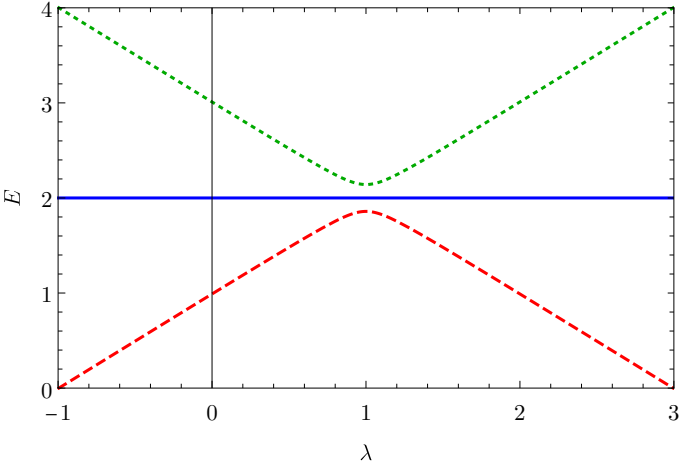}
\end{center}
\caption{Avoided crossings for the three-level model}
\label{fig:AC}
\end{figure}

As an illustrative example we consider a trivial toy problem that we deem to
be quite interesting: the Hamiltonian operator in matrix form%
\begin{equation}
\mathbf{H}(\beta ,\lambda )=\left(
\begin{array}{ccc}
3-\lambda & \beta & 0 \\
\beta & 2 & \beta \\
0 & \beta & 1+\lambda%
\end{array}%
\right).  \label{eq:H_3x3}
\end{equation}%
When $\beta =0$ the three eigenvalues cross at $\lambda =1$ and the three
eigenvectors are degenerate (they are obviously linearly independent).
However, when $\beta \neq 0$ the eigenvalues do not cross for real values of
$\lambda $ and exhibit avoided crossings as shown in Figure~\ref{fig:AC} for
$\beta =0.1$.

The characteristic polynomial is
\begin{equation}
Q(E,\lambda )=\frac{\left( 2-E\right) \left(
50E^{2}-200E-50\lambda ^{2}+100\lambda +149\right) }{50},
\label{eq:Q_3x3}
\end{equation}

so that%
\begin{equation}
Disc_{E}(Q(E,\lambda ))=\frac{\left( 50\lambda ^{2}-100\lambda
+51\right) ^{3}}{31250}.  \label{eq:Disc(Q)_3x3}
\end{equation}%
We appreciate that the three eigenvalues coalesce at any of the two EPS $%
\lambda _{EP}=1+\sqrt{2}i/10$ and $\lambda _{EP}^{\ast }$ that are branch
points of order two. The structure of the avoided crossings in this toy
model is similar to that of the modified Lipkin model for $N=3$\cite{HSG05}.

At $\lambda _{EP}$ (a similar analysis can be carried out at $\lambda
_{EP}^{\ast }$) the matrix $\mathbf{H}$ has only one eigenvalue $E\left(
\lambda _{EP}\right) =2$ and only one eigenvector%
\begin{equation}
\mathbf{v}_{1}=\frac{1}{2}\left(
\begin{array}{c}
-i \\
\sqrt{2} \\
i%
\end{array}%
\right),  \label{eq:v_1}
\end{equation}%
so that $\mathbf{H}$ is defective. By means of the Jordan chain%
\begin{eqnarray}
\left( \mathbf{H}-2\mathbf{I}_{3}\right) \mathbf{v}_{2}
&=&\mathbf{v}_{1},
\nonumber \\
\left( \mathbf{H}-2\mathbf{I}_{3}\right) \mathbf{v}_{3}
&=&\mathbf{v}_{2}, \label{eq:Jordan_chain}
\end{eqnarray}%
where $\mathbf{I}_{3}$ is the $3\times 3$ identity matrix, we obtain two
additional vectors $\mathbf{v}_{2}$ and $\mathbf{v}_{3}$ and the matrix%
\begin{equation}
\mathbf{U}=\left(
\begin{array}{ccc}
\mathbf{v}_{1} & \mathbf{v}_{2} & \mathbf{v}_{3}%
\end{array}%
\right) =\left(
\begin{array}{ccc}
-i/2 & 5\sqrt{2} & 0 \\
1/\sqrt{2} & 5i & 50\sqrt{2} \\
i/2 & 0 & 50i%
\end{array}%
\right),  \label{eq:U_3x3}
\end{equation}%
that converts $\mathbf{H}$ into a Jordan canonical form%
\begin{equation}
\mathbf{U}^{-1}\mathbf{HU}=\left(
\begin{array}{ccc}
2 & 1 & 0 \\
0 & 2 & 1 \\
0 & 0 & 2%
\end{array}%
\right).  \label{eq:Jordan_form}
\end{equation}%
In this simple case we can easily obtain the EPS directly from the
eigenvalues but in most nontrivial problems the use of the discriminant
leads to far simpler expressions.

\section{Three-term recurrence relations}

\label{app:Rec-Rel}

Suppose that there is an orthonormal basis set $\left\{ \left\vert
i\right\rangle ,i=0,1,\ldots \right\} $ such that%
\begin{equation}
H\left\vert i\right\rangle =H_{i,i-1}\left\vert i-1\right\rangle
+H_{i,i}\left\vert i\right\rangle +H_{i,i+1}\left\vert
i+1\right\rangle, \label{eq:three_term_rec_rel_H}
\end{equation}%
where $H_{i,j}=H_{i,j}^{\ast }=H_{j,i}$. Therefore, if we expand%
\[
\psi =\sum_{i}c_{i}\left\vert i\right\rangle,
\]%
the Schr\"{o}dinger equation $H\psi =E\psi $ becomes a three-term recurrence
relation for the coefficients $c_{j}$:%
\begin{eqnarray}
A_{i}c_{i-1}+B_{i}c_{i}+A_{i+1}c_{i+1}
&=&0,\;A_{i}=H_{i-1,i},\;B_{i}=H_{ii}-E,  \nonumber \\
&&i=0,1,2\ldots ,\;c_{-1}=0.  \label{eq:three_term_rec_rel_c_i}
\end{eqnarray}%
One commonly obtains approximate energies by means of the truncation
condition $c_{j}=0$, $j>N$, so that the roots of the characteristic
polynomial given by the secular determinant%
\begin{equation}
D_{N}=\left\vert
\begin{array}{cccccc}
B_{0} & A_{1} & 0 & \cdots & \cdots & 0 \\
A_{1} & B_{1} & A_{2} & 0 & \cdots & 0 \\
\vdots & \vdots & \vdots & \ddots & \cdots & \vdots \\
0 & 0 & \cdots & A_{N-1} & B_{N-1} & A_{N} \\
0 & 0 & \cdots & 0 & A_{N} & B_{N}%
\end{array}%
\right\vert,  \label{eq:D_N}
\end{equation}%
converge from above toward the actual energies of the physical problem when $%
N\rightarrow \infty $. These determinants can be efficiently generated by
means of the three-term recurrence relation\cite{S63,HO91,F01}%
\begin{equation}
D_{N}=B_{N}D_{N-1}-A_{N}^{2}D_{N-2},\;N=0,1,\ldots,
\label{eq:rec_rel_D_N}
\end{equation}%
with the initial conditions $D_{-1}=1$, $D_{j}=0$ for $j<-1$. Notice that
the dimension of the determinant $D_{N}$ is $N+1$.

\setcounter{section}{0}

\begin{figure}[tbp]
\begin{center}
\includegraphics[width=9cm]{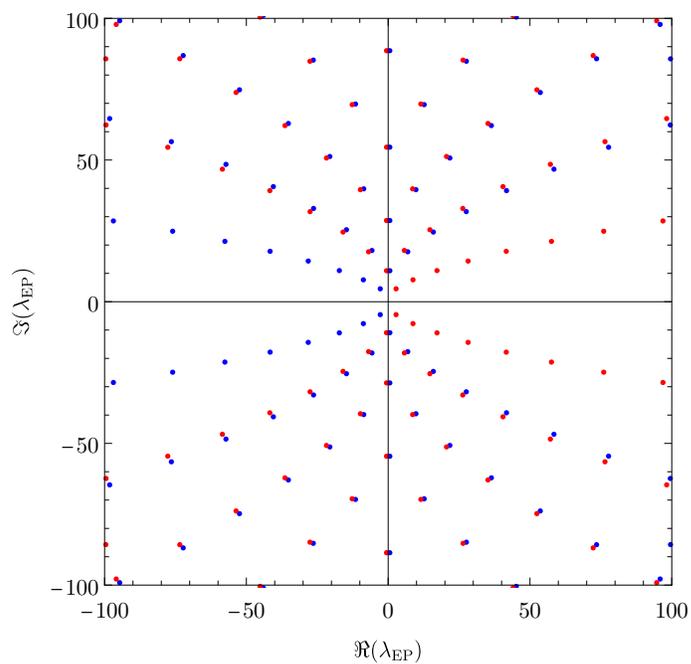}
\end{center}
\caption{EPS for the polar symmetric top in a uniform electric field for $%
(M=1,K=1)$ (blue circles) and $(M=1,K=-1)$ (red circles)}
\label{fig:SYMTOPM1K1M1KM1}
\end{figure}

\end{document}